\documentclass[reprint,amsmath,amssymb,aps,twocolumn]{revtex4-2}
\UseRawInputEncoding
\usepackage{subcaption}
\usepackage{caption}
\usepackage{graphicx}
\usepackage[export]{adjustbox} 

\begin{document}
\title{ Time  irreversibility in Statistical Mechanics }

\author{Dominique Levesque }
\affiliation{ Laboratoire des 2 infinis Ir\`ene Joliot-Curie, Universit\'e Paris-Saclay}
\author{Nicolas Sourlas}
\affiliation{Laboratoire de Physique de l'€™Ecole Normale Sup\'erieure, ENS, Universit\'e PSL, F-75005 Paris, France}

\date{\today}

\begin{abstract}
\noindent 
One of the important questions in statistical mechanics is how irreversibility (time's arrow)
occurs when Newton equations of motion are time reversal invariant.
One objection to irreversibility is based on Poincar\'e's recursion theorem: a classical 
hamiltonian confined system returns after some time, so-called Poincar\'e recurrence time (PRT),
close to its initial configuration. Boltzmann's reply was that for a $N \sim 10^{23} $ macroscopic number of particles,
PRT is very large and exceeds the age of the universe. 
In this paper we compute for the first time, using molecular dynamics, 
a typical recurrence time $ T(N)$ for a realistic case of a gas of  $N$ particles. 
We find that $T(N) \sim N^z \exp (y N) $ and determine the exponents $y$ and $z$ 
for different  values of the particle density and temperature. We also compute $y$ analytically using 
Boltzmann's hypotheses. 
 We find an excellent agreement with the numerical results.  
 This agreement validates Boltzmann's hypotheses  which are not yet mathematically proven. We establish that 
that $T(N) $ exceeds the age of the Universe for a  relatively small 
 number of particles, much smaller than $ 10^{23} $.

\end{abstract}


\maketitle
 
  One of the important questions in statistical mechanics is how time irreversibility 
  occurs ({\it{time's arrow)}}, despite the fact that the equations 
of motion obey time reversal invariance. 
This is a long and still debated issue.
The typical example in the literature is the following.
Dropping a small quantity of coffee in the corner of a glass 
of milk, the coffee diffuses slowly and the liquid becomes homogeneous and 
will stay homogeneous (the state of maximal entropy) as long as we wait.
But according to Poincar\'e's recursion theorem\cite{P} 
a classical hamiltonian confined system, evolving according to the 
 classical equations of motion, returns close to its initial condition after some time $t$. 
 This time  $t$ is called today the Poincar\'e recurrence time (PRT). 
 
Boltzmann's answer to this apparent paradox is that time irreversibility in macroscopic 
physical systems is due to their very large number of particles. 
Starting from any initial condition, a macroscopic system 
will move toward  states of increasing entropy and thermal equilibrium. Boltzmann made the 
ergodic hypothesis and argued that during time evolution the physical system spends a time 
in any subspace of phase space proportional to the volume of this subspace, i.e. the 
exponential of the corresponding Boltzmann entropy. When the number of particles is very 
large the volume of these subspaces is very large and the system spends a very large time 
in them before coming back close to its initial position. 
As a consequence PRT becomes very large and unobservable when the number of 
particles increases.

Boltzmann estimated in this way the PRT for a gas and  argued that for $10^{23}$ 
molecules this time 
is larger than the estimated life of the universe. 
 For an excellent discussion on the subject see the article of Lebowitz\cite{L} 
 and the book of Gallavotti \cite{G}. This Boltzmann's scenario has not be proven 
 mathematically\cite{V}, however the  majority, but not all, of physicists today
  believe that Boltzmann's scenario is true. Some people and in particular 
  the Prigogine school\cite{PR} argue that contrary 
  to Boltzmann's argument, 
  irreversibility is due to elementary interactions which violate time reversal symmetry\cite{PR}. 
  
 To the best of our knowledge the PRT has never being precisely calculated up 
 to now, except for the case of some very particular systems \cite{BDZL}.
 
 In the present paper we estimate $T(N)$, a typical
PRT for a realistic system of $N$ particles interacting by a pair potential. 
 We use molecular dynamics to numerically measure $T(N)$.  
 We estimate $T(N)$ for a relatively small number of particles $N$, interacting via a Lennard-Jones 
 potential and use finite size scaling arguments to extrapolate towards the limit of large $N$.
 We find that  $T(N) \sim N^z \exp (y N) $ and compute the exponents $y$ and $z$ for different 
values of the particle density and the temperature of the gas.

More specifically we consider $N$ particles confined in a sphere by a surface potential:
$v_s(R, r)=A/(|\overrightarrow{R}-\overrightarrow{r}|)^{2n}$ where $\overrightarrow{R}$ 
is the position of a sphere point and $\overrightarrow{r}$ a particle position.
 At time $t=0$ all particles are in a half sphere, i.e. their  coordinates $x$ are $ < 0 $, and 
 their velocity obeys a gaussian distribution corresponding to an equilibrium temperature $T_m$. 
 The particles interact via a Lennard Jones potential type :
$$ v(r_{ij}) = 4 \epsilon ( ( \frac{\sigma}{ r_{ij}})^{12} - 
(\frac{\sigma}{ r_{ij}})^6 ) + \epsilon $$ 
if $r_{ij} < 2^{1/6} \sigma$ and $ v(r_{ij}) = 0$ if $r_{ij} > 2^{1/6} \sigma$.
$\sigma$ and $\epsilon $ are chosen respectively as the length and energy units. 
We integrate the Newton equations of motion by using a variant of the Verlet algorithm, 
described in reference \cite{LV}, which, based on integer arithmetic, 
strictly respects the time reversal symmetry. Indeed, it has been shown that rounding-off errors in floating-point        
arithmetic can induce violations of the time reversal symmetry. A strict respect of  
time reversal symmetry allows us to exclude the argument that irreversibility is due to 
interactions violating time reversal symmetry \cite{PR}.
\begin{figure}[h]
\includegraphics[height=5.6cm, width=7.5cm]{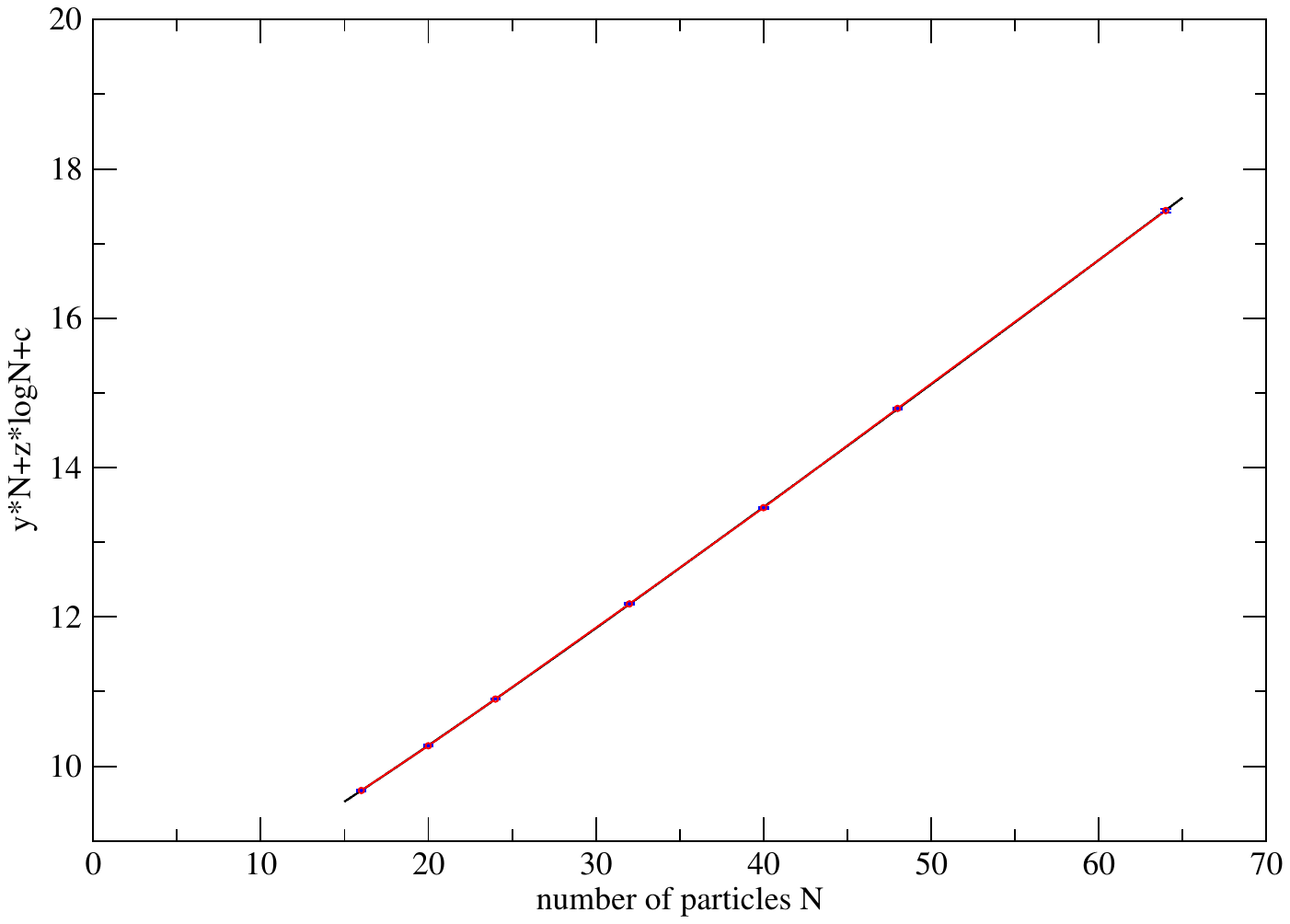}  
\caption{Typical fit of $\log(T(N))$ at density $\rho$=0.064 and $T_m=1.6$}
\label{Fig. 1} 
\end{figure}

 Even for small $N$ systems it would take a too long a time in order for the $N$ particles to 
 come back almost exactly to their velocities and positions in the initial configuration. 
 Instead we consider $n_i(N)$ systems differing by the initial 
 conditions and define for each system a recurrence time $T_i(N)$ which is chosen
as the time it takes, for the i'th system, so that $ 3 N / 4 $ particles (or more) 
 come back to the same half sphere as when $ t=0 $, i.e the time
separating the occurence of an important density fluctuation
in  a half sphere which, usually in equilibrium states, contains $\sim N/2$ particles.

Increasing $N$, $T_m$ being constant, we increase the radius of the sphere to 
keep the density of particles $\rho$ constant. We observe that $T_i(N)$ 
strongly fluctuates. For every $N$, we average $T_i(N)$ 
over a large number $n_i(N)$ of initial conditions and 
compute $T(N)$, the $T_i(N)$ average over those initial conditions. 
For $ N < 25 $ we simulate  $n_i(N)=10000 $ 
samples, while for the largest number of $N$ we simulate, which depends on the density $\rho$, 
$n_i(N) \sim 1000$. 
We found that the main dependence of $T(N)$ on $N$ is exponential and is well described by the formula 
$T(N) = c^{'} N^z \exp(y N) $ or $ \log (T(N)) = yN +z\log(N) +c $ as it is shown in Fig. 1.
\begin{figure}[h]
\includegraphics[height=5.6cm, width=7.5cm]{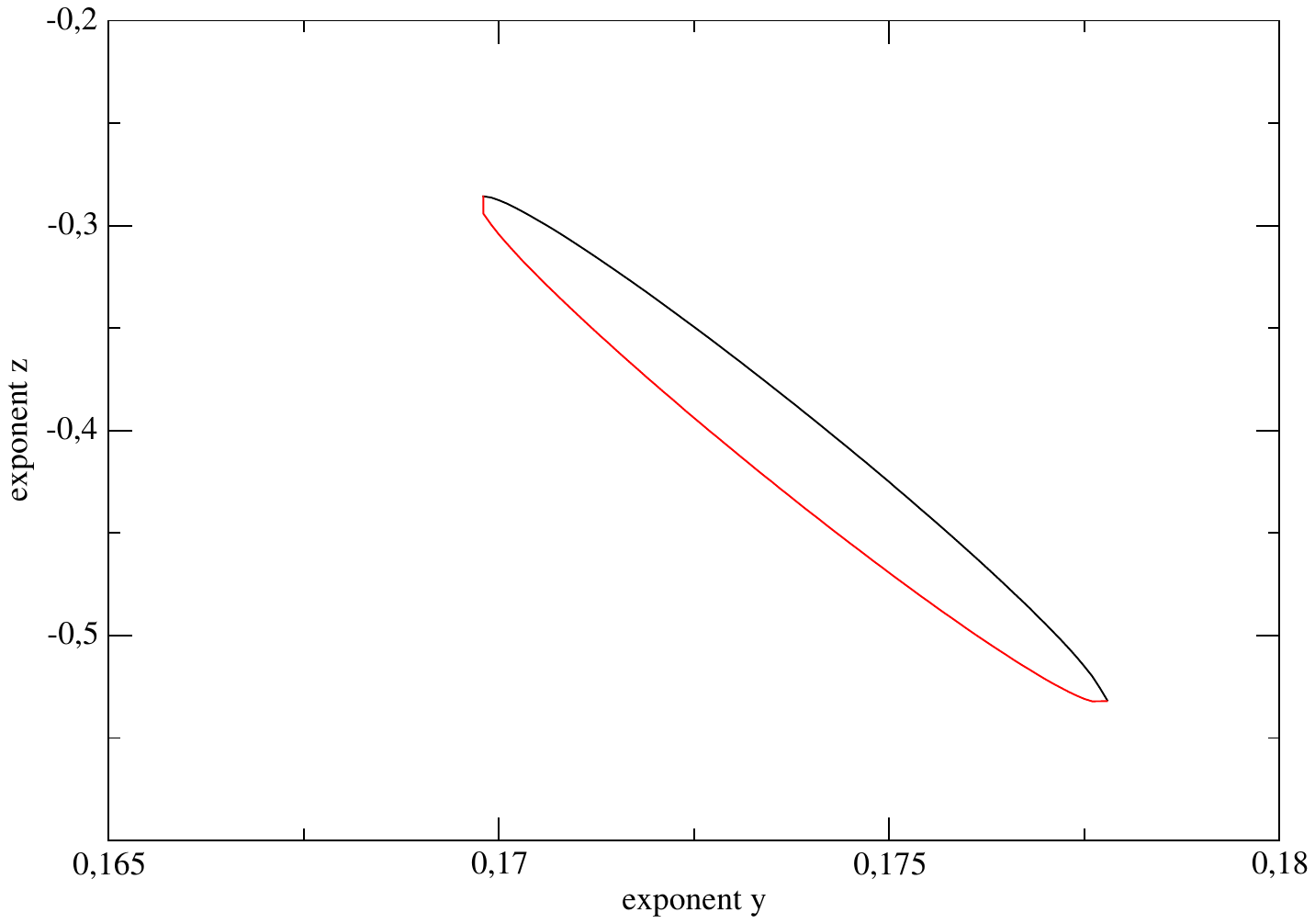}  
\caption{ Region of $y$ and $z$ exponents  with 90\% confidence level, $\rho$=0.064 and $T_m$=1.6}
\label{Fig. 2} 
\end{figure}
We compute the exponents $y$ and $z$.  $y$ and $z$ are correlated. 
This is illustrated in Fig. 2 which shows their allowed values  with 90 percent 
confidence.  $y$ and $z$ depend on the density $\rho$ and $T_m$. The results
of these computations are shown in Table I.

 In the first column is listed the particle density $\rho$, in the second the range of the 
exponents $y$ and $z$ with  90 percent  confidence and $ T_m=1. 6$ (cf. Fig. 2), 
in the third the same  for $T_m=10. 0$, in the last column $N_{m}$ is the maximum 
number of particles we simulated for that density. The quality of the fit is excellent, 
as shown in Fig 1, with $\chi^2$ per degree of freedom $\chi^2 < 1.$  
The exponent $y$ increases with the particle density $\rho$. The dependance on $T_m$ 
is weaker.

\begin{table}
\captionsetup{font=scriptsize}
\begin{tabular}{|l|c|c|r|}
\hline
$ \ \ \rho  $  &  $ T_m=1.6$              &   $ T_m=10.0 $           &  $N_{m}$  \\
\hline            
$ 0.064 $ & $ 0.170 < y < 0.178 $ &  $ 0.163 < y < 0.173 $  &  $64$     \\
\hline            
$ 0.064 $ & $ -0.526 < z < -0.286 $  &   $ -0.769 < z < -0.437 $  &  $64$     \\
\hline
$ 0.127 $ & $0 .218 < y < 0.244 $  &  $ 0.201 < y < 0.227  $  &  $ 48 $   \\
\hline            
$ 0.127$  & $-0.916 < z < -0.221 $  &  $ -1.112 < z < -0.447  $  &  $48$    \\
\hline
$ 0.191 $ & $ 0.281 < y < 0.312 $  &  $0.233 < y < 0.255 $  &  $ 48 $   \\
\hline            
$ 0.191 $ & $ -0.938 < z < -0.175 $    &   $ -0.748  < z < -0.242 $  &  $ 48 $    \\
\hline
\end{tabular}
\caption{$y$ and $z$ values obtained from fits of  $\ln(T(N))$.}
\label{Tab.1}
\end{table}

Clearly, our definition is different from the  PRT standard definition which requires 
recurrence in the phase space, i.e. both for the particle position and momentum. 
$T(N)$ is only the recurrence time of  an important
density fluctuation. Nevertheless the algorithm conserves energy and we have verified 
that the particle velocity distribution stays gaussian, while individual positions and 
velocities of the particles may differ from their initial value. Hence $T(N)$ corresponds
to the recurrence of configurations much less  precisely characterized than for PRT, so $T(N)$
should be certainly an underestimation of PRT.

We can also compute $y$ assuming Boltzmann's hypotheses to be true. We use the same parameters 
$\rho$ and $T_m$ for the gas as in the numerical method. 
We  assume that the system is ergodic and that 
all the system configurations are equiprobable, during time evolution.
The  system should stay in any fraction of its space phase a time proportional  to the volume 
of this fraction of phase space, i.e. 
proportional to the exponential of the fraction entropy. According to this hypothesis, $T(N)$ would be 
approximatively proportional to the  ratio of the system entropy exponentials $\exp(S_{eq}-S_f)$  in the two considered 
configurations, whith $S_{eq}$  the entropy of  the equilibrium configuration with $\sim N/2 $  particles in each half sphere, 
and $S_f$ the entropy of the non-equilibrium configuration where $\sim 3N/4 $ and  $\sim N/4 $ particles 
are respectively in each half sphere.
Neglecting the interaction between particles and using the
 ideal gas entropy,  leads to  $T(N) \sim \exp (y N) $  with $y_{id}=0.13$. 
This rough estimate can be improved by taking into account excluding volume effects.
Excluding volume effects reduce the number of allowed configurations in any subspace of 
phase space and therefore the corresponding entropy. They depend on the diameter 
of the particles $d_s$. 
A system interacting by a short-ranged repulsive pair potential
is similar to a system of  hard spheres with an effective diameter\cite{Ro} $d_s$ defined by 
$$ d_s = \int \Bigr[ 1-\exp({-v(r)/k_BT_m)} \Bigl] \, dr $$ 
where $k_B$ is the Boltzmann constant, and $T_m$ the average equilibrium temperature. 
 The excess free energy\cite{Ca} $f_{e}$ per
particle of the effective hard sphere system, is given for the considered densities 
at a very good approximation by
 $$ f_{e}(\rho)=Nk_BT_m  \frac{4 \eta-3\eta^2}{(1-\eta)^3} $$  with $\eta=\pi \rho d^3_s/6$
 and, and so the excess entropy per particle by $S_{e}(\rho)=-f_{e}(\rho)/(Nk_BT_m)$, since a hard sphere system 
 has no potential energy. Then $y$ is estimated by
$$y=S_{e}(\rho/2)/4+3 S_{e}(3\rho/2)/4-S_{e}(\rho)+y_{id}$$
The results of these estimations ($y_{ana} $) together with the numerical 
estimations ($y_{num}$) are given in Table II. 

\begin{table}
\captionsetup{font=scriptsize}
\begin{tabular}{|l|c|c|r|}
\hline
$ \ \ \rho  $     &   $T_m $    &       $y_{num}$              & $y_{ana} $   \\
$ \ \  0.064 $    &   $ 1.6 $   & $ 0.170 < y=0.174 < 0.178$   & $0.171 $   \\
$ \ \  0.064 $    &   $ 10.0.$ & $0.163 < y=0.168 < 0.173$    & $0.160 $   \\
$ \ \  0.127 $    &   $ 1.6. $ & $ 0.218 < y=0.231 < 0.244$   & $0.230$   \\
$ \ \  0.127 $    &   $ 10.0 $ & $0.201 < y=0.214 < 0.227$    & $0.198 $   \\
$ \ \  0.191 $    &   $ 1.6 $  & $0.281 < y=0.295 < 0.312 $   & $0.318 $   \\
$ \ \  0.191 $    &   $ 10.0 $ & $0.233 < y=0.247 < 0.255 $   & $0.251 $   \\
\hline
\end{tabular}
\caption{values of the exponent  $y$, $y_{num}$ computed numerically and $y_{ana} $ 
computed analytically for the different values of $\rho$ and $T_m  $. }
\label{Tab.2}
\end{table}

The quantitative agreement of these $y$ estimations made by two completely different methods  
is quite remarkable. 
It  strongly confirms the validity of the assumptions entering the analytic 
estimation of $y$. 

Note that in the previous analytic estimation the exact initial and final states 
of the particles do 
not play any role, only their relative entropy matters. In our simulations all the $N$ 
particles were at time $t=0$ located in the half sphere where the coordinate $ x < 0 $ and 
in average at $ t= T(N) $ $3N/4$ particles were at the same half sphere. We also considered the case 
of initial condition where the coordinate $ x < 0 $ and compute the average time $ t=T^{'}(N) $ 
when $3N/4$ particles are in a different half sphere, the half sphere where $ y < 0 $. 
Obviously the entropies of the final configurations in the two cases are the same.
We found $ T(N) =  T^{'}(N)$, confirming Boltzmann's  hypothesis that only the relative entropy matters.

Assuming the validity of our estimation method, we can calculate $y$ for higher densities.
 For $\rho= 0.2 $ and $T_m=1.6$ we find $y=0.34 $ while for 
$\rho= 0.4 $ and $T_m=1.6$,  $y= 1.02$. We can also compute the average time $ t=T^{''}(N) $ 
when $9N/10$ (instead of $3N/4$) particles come back to the initial half sphere. 
For $\rho= 0.2 $ and $T_m=1.6$ we find $ y= 1.2 $.

It is clear that because of its exponential increase with the number 
of particles, $T(N)$ will become larger than the life of the universe provided that the number 
of particles is large enough, say for $N > N_u$. 
The current estimate of the life of the universe is 
$\sim\,14 \times 10^9$ years or $\sim 45 \times 10^{16}$ seconds. 
The time $ T(N) $ we compute measures the number of steps of our integration algorithm. 
In order to compare with the age of the universe we have to convert it into seconds. 
For this we need the values of the 
parameters $ \epsilon $ and $\sigma $ entering the definition of the Lennard Jones potential, 
the mass $m$ of the particles and the value of the discrete time steps $h$ taken for integrating 
the equations of motion, expressed in the time unit $\tau=\sqrt{m \sigma^2/\epsilon}$. 
For every value of the density and the temperature, $h$ is kept constant, 
independent of the number of particles $N$.
Assuming  for $\epsilon$ and $\sigma$, the typical values of a rare gas interaction, 
for instance argon\cite{ar}, 
we can estimate $N_u$.  $N_u$ decreases when the gas density increases. 
We find that for $\rho= 0.064 $ $N_u \sim 294 $ while for $ \rho=0 .191 $ $N_u \sim 171 $.   

The surprising strong agreement of our simulations with the 
analytic estimation of 
the exponent $y$, strongly supports the validity of the assumptions on which the 
estimation is based. 

In conclusion, our results overwhelmingly support Boltzmann's point of view on the origin of 
irreversibility in 
statistical mechanics and there is no need for interactions violating time reversal symmetry. 

We are very thankful to Vincent Hakim for his help  and his suggestion to perform the analytic estimation of the exponent $y$.

\end{document}